# Near-infrared flares from accreting gas around the supermassive black hole at the Galactic Centre

R. Genzel[1,2], R. Schödel[1], T. Ott[1], A. Eckart[3], T. Alexander[4], F. Lacombe[5], D. Rouan[5] & B. Aschenbach[1]

[1]Max-Planck-Institut für extraterrestrische Physik, Giessenbachstr. 1, 85748 Garching, Germany
[2]Department of Physics, Le Conte Hall, University of California, Berkeley, California 94720, USA
[3]I. Physikalisches Institut, Universität zu Köln, Zülpicher Strasse 77, 50937 Köln, Germany
[4]Faculty of Physics, The Weizmann Institute of Science, PO Box 26, Rehovot 76100, Israel
[5]Observatoire de Paris—Section de Meudon, 5 Place Jules Janssen, 92195 Meudon Cédex, France

Recent measurements of stellar orbits[1–3] provide compelling evidence that the compact radio source Sagittarius A* (refs 4, 5) at the Galactic Centre is a 3.6-million-solar-mass black hole. Sgr A* is remarkably faint in all wavebands other than the radio region[6,7], however, which challenges current theories of matter accretion and radiation surrounding black holes[8]. The black hole's rotation rate is not known, and therefore neither is the structure of space-time around it. Here we report high-resolution infrared observations of Sgr A* that reveal 'quiescent' emission and several flares. The infrared emission originates from within a few milliarcseconds of the black hole, and traces very energetic electrons or moderately hot gas within the innermost accretion region. Two flares exhibit a 17-minute quasi-periodic variability. If the periodicity arises from relativistic modulation of orbiting gas, the emission must come from just outside the event horizon, and the black hole must be rotating at about half of the maximum possible rate.

During observations of the Galactic Centre with the new Very Large Telescope adaptive optics (AO) imager NACO[9,10] on 9 May 2003, we observed a powerful flaring by a factor of 5 in the H-band (1.65-µm) emission towards SgrA* (Fig. 1, Table 1). The flare lasted for 30 min. Its rise and decay can be well fitted by an exponential of timescale 5 min. In a second observing run in June 2003, we observed two more flares on two consecutive days, this time in the $K_s$ band (2.16µm). The K-band flares rose to a factor of 3 above the quiescent level, and each lasted for ,85 min. Their characteristic rise/decay times were 2 to 5 min. Both flares exhibited significant and similar temporal substructure (Fig. 2). The 16 June flare showed five major peaks spaced by 13 to 17 min, resembling a 15–40%, quasi-periodic modulation of the overall flare profile. The 15 June flare had three major peaks separated by 14 and 17 min, followed by a weaker peak 28 min later.

The power spectrum analysis for both flares (Fig. 2) exhibits a significant peak with a time period of $16.8 \pm 2$ min, thus confirming the reality of the structure seen directly in the light curves. The comparison star S1 clearly does not show such a quasi-periodic structure. However, the question arises whether this periodicity is truly a fundamental and significant property of all SgrA* flares[11], or whether it is caused by fluctuations in a 'red noise' power spectrum. More data will tell, but the fact that two events separated by about 93 periods show similar substructure is very suggestive. Because of the lack of continuous time coverage, we cannot make a statement on the substructure of the May H-band flare. Finally we found a fourth flare in re-analysing earlier archival $L_0$-band (3.76-µm) NACO data taken on 30 August 2002. This flare rose to 70% above the quiescent emission (Fig. 2), and had a decay time of ,10 min. The infrared flares all originated from within a few milliarcseconds, or a few hundred Schwarzschild radii, of the black-hole position (Table 1). That position was determined from the focus of the best-fitting Kepler orbit of the star S2[1,3].

Before and after the strong flares and in further images taken in H, $K_s$ and $L_0$ in August 2002 and in March, May, June and July 2003, we also detected a 'quiescent' source towards the position of SgrA* (Table 1; see also Y. Clenet, D.R., D. Gratadour, E. Gendron, F.L., A. M. Lagrange, G. Rousset, R.G. & R.S., manuscript in preparation). The source is spatially unresolved, and exhibits small-amplitude variability (20–60%) on a scale of 20–30 min. It is not clear whether this variability also has some kind of periodicity. The H-band observations of 16 June may have several peaks spaced by ,28 min. Similar spacings are seen in the off-flare structure on 15 and 16 June (Fig. 2). Hour-to-week-scale variability of the SgrA* $L_0$-band emission was also discovered in May and June 2003 at the Keck telescope[12]. It might thus also be appropriate to interpret the SgrA* infrared emission as exhibiting a range of variability amplitudes and timescales of which the observed flares are just the most extreme at present. Wire-grid K-band polarimetry taken on 14 June 2003 shows that the 'quiescent' source is significantly more polarized than the average of stars in its immediate vicinity. More calibration needs to be carried out before a precise value and position angle can be quoted. The quiescent source is coincident with the position of the black hole to within 10–20 mas, where the accuracy is limited by its faintness and proximity to S2. It is highly unlikely that the source is a star projected along the line of sight towards SgrA*, given that it is variable, polarized and coincident with SgrA* in four epochs between March and July. For Poisson statistics and four detected flares over a total period of 25 h we estimate a rate between 2 and 6 flares per day, at least twice the rate of X-ray flares from Chandra monitoring in 2002 ($1.2 \pm 0.4$ flares per day; ref. 13). This high flaring rate, along with the complex time substructure of the two K-band flares, rules out the possibility that the flares are due to gravitational microlensing of cusp stars by the black hole. The lensing rate is predicted to be several orders of magnitude smaller than the observed flaring rate[14]. It is unclear whether our high flare rate is consistent with the previous limits on the infrared activity of SgrA*[7]. Most of the previous studies relied on speckle imaging, for which integration times of >=10 min (a significant fraction of the flare duration) were required to confidently detect a source at the measured flux densities. A retroactive analysis of some of our best data in the mid-1990s gives an inconclusive answer as to whether such variability was or was not present. SgrA* may be more active at present than in the past decade.

Figure 3 summarizes the radio to X-ray spectral energy distribution (SED) of the emission from SgrA*, for both states. The quiescent infrared SED is red, with flux densities decreasing with frequency. Taking into account the variation in 'quiescent' $L_0$-band flux from 2002 to 2003 (Table 1), the inferred 1.6–3.8μm SED has a power-law spectral index of 22.1 (±1.4). As predicted by a number of models[15–18], the 'quiescent' infrared flux densities lie approximately on the extrapolation of the millimetre/submillimetre synchrotron emission to high energies, in accordance with a standard power-law synchrotron SED. Our detection of polarization is also consistent with the synchrotron model. Models with only a thermal population of electrons[15,16] underpredict the infrared emission, whereas models with an additional non-thermal, power-law component of energetic electrons ($\gamma_e = E_e/m_e c^2 >= 10^{2.5}$, where $E_e$ and $m_e$ are the energy and mass of the electron; ref. 17) come closer to, but are still below, the observed emission. The observed infrared and X-ray flare durations, rise/decay times and band luminosities are similar (Fig. 3; refs 13, 19, 20). The higher fractional flare amplitudes in X-rays are probably just a reflection of the much fainter quiescent emission. Although the millimetre/submillimetre emission of SgrA* is also variable, almost all of these variations are on timescales of several days to a few hundred days[21,22]. One exception is a one-hour-duration, 30% amplitude event seen in March 2000 at a wavelength of 2mm (ref. 22). Although we do not have a simultaneous SED of the infrared flares, our data suggest that they may be bluer than the quiescent emission with a flux density that is approximately constant as a function of frequency (Fig. 3). The infrared flares may be synchrotron emission as well, if turbulence, magnetic reconnection or shocks are effective in accelerating for short periods a significant fraction of the energetic electrons to $\gamma_e >= 10^3$ (Fig. 3; ref. 17). In that case, the increased emission is due to an acceleration event, similar to a solar flare, rather than enhanced accretion[17,18]. Models with suitably upscaled fractions of very energetic electrons may account for the luminosities/fluxes of the infrared flares (as well as the 'quiescent' emission).

However, if the observed flare fluxes do represent the intrinsic SED of the flare, an alternative emission mechanism is required. In that case, the infrared flares may be thermal bremsstrahlung or black-body radiation from a second component of moderately hot gas (temperature in excess of a few times $10^3$ K) associated with individual accretion events of very dense gas, of total energy release $>=10^{39.5}$ erg (for an assumed radiation efficiency of ~10%, this energy requires an accreted mass of a few times $10^{19}$ g, comparable to that of a comet or a small asteroid), while the observed radio/submillimetre emission in that case would come from a jet or be optically very thick. A key future test of the two alternatives will be measurements of the simultaneous infrared flare SED and polarization. The flares' location close to the central black hole, as well as the temporal substructure, poses a serious challenge to models in which the flares originate from rapid shock cooling of a large-scale jet, or are due to passages of stars through a central accretion disk[23]. The few-minute rise and decay times, as well as the quasiperiodicity, strongly suggest that the infrared flares originate in the innermost accretion zone, on a scale less than ten Schwarzschild radii (the light travel time across the Schwarzschild radius of a 3.6-million-solar-mass black hole ($1.06 \times 10^{12}$ cm) is 35 s). If the substructure is a fundamental property of the flow, the most likely interpretation of the periodicity is the relativistic modulation of the emission of gas orbiting in a prograde disk just outside the last stable orbit (LSO)[24]. If the 17-min period can be identified with this fundamental orbital frequency, the inevitable conclusion is that the Galactic Centre black hole must have significant spin. The LSO frequency of a 3.6-million-solar-mass, non-rotating (Schwarzschild) black hole is 27 min. Because the prograde LSO is closer in for a rotating (Kerr) black hole, the observed period can be matched if the spin parameter is $J/(GM_{BH}/c)$ ¼ 0.52 (±0.1, ±0.08, ±0.08, where J is the angular momentum of the black hole); this is half the maximum value for a Kerr black hole[25,26]. (The error estimates here reflect the uncertainties in the period, black-hole mass ($M_{BH}$) and distance to the Galactic Centre[3], respectively; G is the gravitational constant.) For that spin parameter, the last stable orbit is at a radius of $2.2 \times 10^{12}$ cm. Recent numerical simulations of Kerr accretion disks indicate that the in-spiralling gas radiates most efficiently just outside the innermost stable orbit[27]. Our estimate of the spin parameter is thus a lower bound.

Other possible periodicities, such as acoustic waves in a thin disk[28], Lense-Thirring or orbital node precession are too slow for explaining the observed frequencies for any spin parameter[25]. (The 28-min timescale of the quiescent emission corresponds to a radius of $3.2 \times 10^{12}$ cm for a prograde orbit of $J/(GM_{BH}/c)$ ¼ 0.52; the last stable retrograde orbit for that spin parameter has a period of 38 min at a radius of $4 \times 10^{12}$ cm). Lense-Thirring precession and viscous (magnetic) torques will gradually force the accreting gas into the black hole's equatorial plane[29]. Recent numerical simulations indicate that a (prograde) disk analysis is appropriate to first order even for the hot accretion flow at the Galactic Centre[27].

1. Schödel, R. et al. A star in a 15.2 year orbit around the supermassive black hole at the centre of the Milky Way. Nature 419, 694–696 (2002).
2. Ghez, A. M. et al. The first measurement of spectral lines in a short-period star bound to the Galaxy's central black hole: A paradox of youth. Astrophys. J. 586, L127–L131 (2003).
3. Eisenhauer, F. et al. Ageometric determination of the distance to the Galactic Center. Astrophys. J. Lett. (in the press); preprint at khttp://arXiv.org/astro-ph/0306220l (2003).
4. Doeleman, S. S. et al. Structure of SgrA* at 86 GHz using VLBI closure quantities. Astron. J. 121, 2610–2617 (2001).
5. Backer, D. C. & Sramek, R. A. Proper motion of the compact, nonthermal radio source in the Galactic Center, SgrA*. Astrophys. J. 524, 805–815 (1999).
6. Baganoff, F. K. et al. Chandra X-ray spectroscopic imaging of SgrA* and the central parsec of the Galaxy. Astrophys. J. 591, 891–915 (2003).
7. Hornstein, S. D. et al. Limits on the short-term variability of SgrA* in the near-IR. Astrophys. J. 577, L9–L13 (2002).
8. Melia, F. & Falcke, H. The supermassive black hole at the Galactic Center. Annu. Rev. Astron. Astrophys. 39, 309–352 (2001).

9. Lenzen, R., Hofmann, R., Bizenberger, P. & Tusche, A. CONICA: The high-resolution near-infrared camera for the ESO VLT. Proc. SPIE 3354 (IR Astronomical Instrumentation), 606–614 (1998).
10. Rousset, G. et al. Design of the Nasmyth adaptive optics system (NAOS) of the VLT. Proc. SPIE 3353 (Adaptive Optics Technology), 508–516 (1998).
11. Benlloch, S., Wilms, J., Edelson, R., Raqoob, T. & Staubert, T. Quasi-periodic oscillation in Seyfert galaxies: Significance levels. The case of Mrk 766. Astrophys. J. 562, L121–L124 (2001).
12. Ghez, A. M. et al. Variable infrared emission from the supermassive black hole at the center of the Milky Way. Astrophys. J. Lett. (submitted); preprint at khttp://arXiv.org/astro-ph/0309076l (2003).
13. Baganoff, F. K.Multi-wavelength monitoring of SgrA* during Chandra observations of multiple X-ray flares. High Energy Astrophysics Division (HEAD) AAS Abstr. 3.02, 35 (2003).
14. Alexander, T. & Sternberg, A. Near-IR microlensing of stars by the supermassive black hole in the Galactic Center. Astrophys. J. 520, 137–148 (1999).
15. Yuan, F., Markoff, S. & Falcke, H. A jet-ADAF model for SgrA*. Astron. Astrophys. 854, 854–863 (2002).
16. Liu, S. & Melia, F. New constraints on the nature of the radio emission in SgrA*. Astrophys. J. 561, L77–L80 (2001).
17. Yuan, F., Quataert, E. & Narayan, R. Nonthermal electrons in radiatively inefficient flow models of SgrA*. Astrophys. J. (submitted); preprint at khttp://arXiv.org/astro-ph/0304125l (2003).
18. Markoff, S., Falcke, H., Yuan, F. & Biermann, P. L. The nature of the 10ksec X-ray flare in SgrA*. Astron. Astrophys. 379, L13–L16 (2001).
19. Baganoff, F. K. et al. Rapid X-ray flaring from the direction of the supermassive black hole at the Galactic Centre. Nature 413, 45–48 (2001).
20. Porquet, D. et al. XMM-Newton observation of the brightest X-ray flare detected so far from SgrA*. Astron. Astrophys. 407, L17–L20 (2003).
21. Zhao, J.-H. et al. Variability of SgrA*: Flares at 1 mm. Astrophys. J. 586, L29–L32 (2003).
22. Miyazaki, A., Tstsumi, T. & Tsuboi, M. Flares of SgrA* at short submm wavelengths. Astron. Nachr. 324, 3–9 (2003).
23. Nayakshin, S., Cuadra, J. & Sunyaev, R. X-ray flares from SgrA*: Star-disk interactions? Astron. Astrophys. (in the press); preprint at khttp://arXiv.org/astro-ph/0304126l (2003).
24. Hollywood, J. M. & Melia, F. General relativistic effects on the infrared spectrum of thin accretion disks in active galactic nuclei: Application to SgrA*. Astrophys. J. Suppl. 112, 423–455 (1997).
25. Bardeen, J. M., Press, W. M. & Teukolsky, S. A. Rotating black holes: Locally non-rotating frames, energy extraction and scalar synchrotron radiation. Astrophys. J. 178, 347–369 (1972).
26. Melia, F., Bromley, C., Liu, S. &Walker, C. K.Measuring the black hole spin in SgrA*. Astrophys. J. 554, L37–L40 (2001).
27. De Villiers, J.-P., Hawley, J. F. & Krolik, J. H. Magnetically driven accretion flows in the Kerr metric I: Models and overall structure. Astrophys. J. (submitted); preprint at khttp://arXiv.org/astro-ph/0307260l (2003).
28. Nowak, M. A., Wagoner, R. V., Begelman, M. C. & Lehr, D. E. The 67 Hz feature in the black hole candidate GRS1915 þ 105 as a possible diskoseismic mode. Astrophys. J. 477, L91–L94 (1997).
29. Bardeen, J.M. & Pettersen, J. A. The Lense-Thirring effect and accretion disks around Kerr black holes. Astrophys. J. 105, L65–L67 (1975).

Acknowledgements This Letter is based on observations at the VLTof the EuropeanObservatory (ESO) in Chile. We thank the teams who developed and constructed the near-infrared camera CONICA and the AO system NAOS, and especially their principal investigators, R. Lenzen, R. Hofmann and G. Rousset.We thank H. Falcke and S. Markoff for access to their database of the SgrA* SED, as well as discussions of emission processes. We are grateful to D. Porquet and P. Predehl for discussions of their XMM data, S. Nayakshin, M. Rees, R. Sunyaev and especially E. Quataert for discussions of accretion disk physics, and A. Sternberg for suggestions on the paper.

Table 1 Photometry and astrometry of the near-infrared emission from SgrA*

| Band | Date | ΔRA* (mas) | ΔDec.* (mas) | $S_\nu$† (mJy) | Duration‡ (min) | Variability§ | Period (min) |
|---|---|---|---|---|---|---|---|
| H quiescent | 2002.63 to 2003.53 | 0 (8) | 23 (8) | 2.8 (0.6) | | ≤0.2 to 0.6 | (28)# |
| $K_s$ quiescent | 2002.63 to 2003.45 | −6 (8) | 15 (12) | 2.7 (0.6) | | 0.3 | (25–30)# |
| L' quiescent | 2002.66 to 2003.45 | −9 (15) | −4 (20) | 6.4 (1.9)‖ 2003.21/.35 17.5 (5)‖¶ 2002.63/.66 | | >2 between 2002 and 2003 | None |
| H flare | 2003.353 | −1.4 (3) | −0.2 (3) | 13 (3) | 30 | 4.7 | ? |
| $K_s$ flare | 2003.455 | −2.5 (4) | 3.4 (4) | 10.5 (3) | 80 | 3.1 | 16.6 |
| $K_s$ flare | 2003.457 | −6.4 (4) | 2.5 (4) | 7.3 (3) | 85 | 3.2 | 17.1 |
| L' flare | 2002.66 | 0 (30) | 0 (30) | 12.6 (4)‖¶ | ≥15 | 0.7 | None |

Numbers in parentheses are (±) 1σ errors in all cases. Dates include fractions of the year.
* Relative to the best fit focus position of S2 orbit[3].
† Dereddened flux density, corrected for A(H) = 4.3, A(K) = 2.8 and A(L') = 1.8. In the case of flares, the quiescent emission is subtracted. Errors contain statistical, systematic uncertainties and variability.
‡ Full-width zero power (FWZP).
§ Excess of variable emission relative to steady emission.
‖ See Y. Clenet, D.R., D. Gratadour, E. Gendron, F.L., A. M. Lagrange, G. Rousset, R.G. & R.S. (manuscript in preparation) for details and calibration of L'-data.
¶ In 2002.6 S2 and SgrA* could not be resolved in L'; for this reason the L' flux densities given are after subtraction of 8.8mJy for S2.
# Very uncertain.

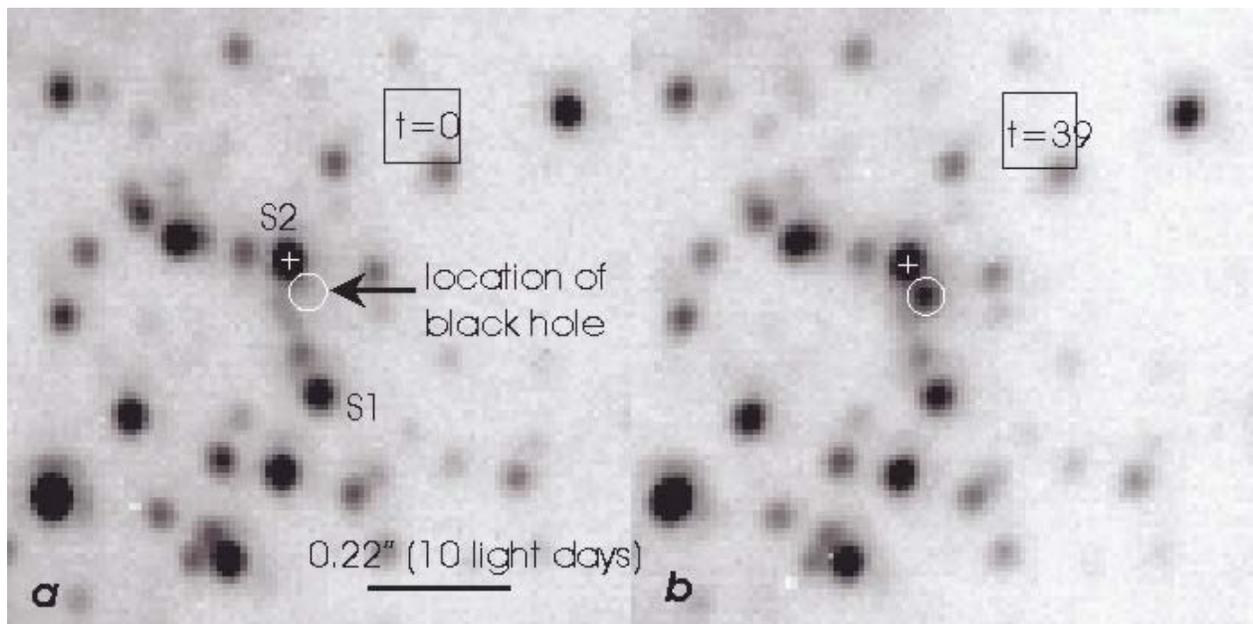

Figure 1 Detection of variable near-infrared emission from SgrA*. Raw H-band (1.65-µm) AO images (40 mas full-width at half-maximum, FWHM) of the central ,1'' of the Milky

Way, obtained with the NACO AO imager[9,10] on UT4 (Yepun) of the ESO VLT, before and
during the H-band flare on 9 May 2003. The image scale is linear. The integration time for
each image was 60 s, from six 10-s individual exposures. The time (in minutes from the
beginning of the set at 6 h 59 min 24 s (UT) on 9 May 2003) is shown in the box in the
upper right of each image. The images were sky-subtracted, flat-fielded and corrected for
bad pixels. North is up and east to the left, scales are for an assumed distance of 7.94 kpc
(25,880 light years)3. The unique infrared wavefront sensor was used to close the loop of
the AO system on the bright supergiant IRS7, ,5.5" north of SgrA*. The fraction of the
power in the diffraction-limited core (Strehl ratio) is about 50% (visible seeing 0.45"
FWHM). The position of the 15-yr-orbit star S2[1–3] is marked by a cross, and the
astrometric location of the black hole is marked by a circle.

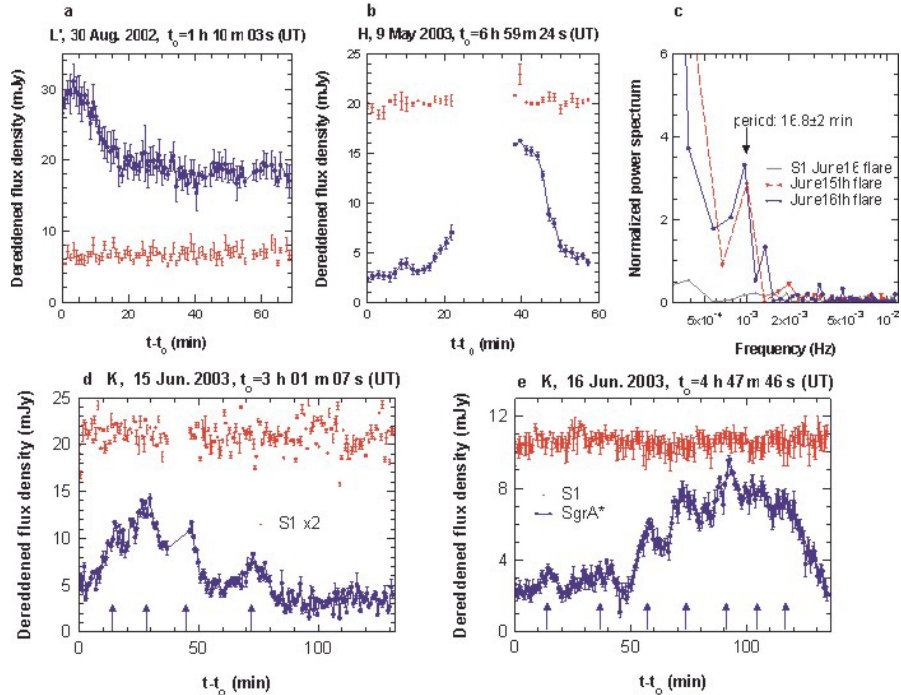

Figure 2 Light curves of the SgrA* infrared flares and quiescent emission in 2002–03.
Flux densities were extracted from the Lucy deconvolved and beam restored images
with two aperture sizes. Error bars (±1j) indicate the combined statistical and
systematic uncertainties. SgrA* data are shown as filled blue circles (connected with a
solid curve). For comparison, the light curves of the nearby star S1 are shown as light red
crosses (0.2" southwest of SgrA*, left panel of Fig. 1). S1 has flux densities comparable to
the SgrA* flare state. In all cases the times are relative to the UT time listed above
each graph. Arrows in d and e mark the substructure peaks discussed in the text.
a, Light curve of the 30 August $L_0$-band flare. As S2 and SgrA* cannot be spatially
separated in this epoch, we have subtracted 8.8 mJy to account for the contribution of S2.
b, Light curve of the 9 May H-band flare. The gap in the data between $t - t_0 = 23$ and
37 min was due to sky observations during that time. c, Power spectra of the flares in
panels d and e, and S1, normalized by their high frequency noise. d, $K_s$-band light curve
on 15 June 2003. Between $t - t_0 = 37$ and 46 min, the AO system was not operational.
Because of our choice of the dichroic for the AO system, the signal-to-noise ratio of this
flare is not as good as in the 16 June flare. For better presentation, the flux densities of S1
were multiplied by a factor of 2. e, $K_s$-band light curve on 16 June 2003. The time
structure of this flare may have the tendency to chirp (periods decrease with time).

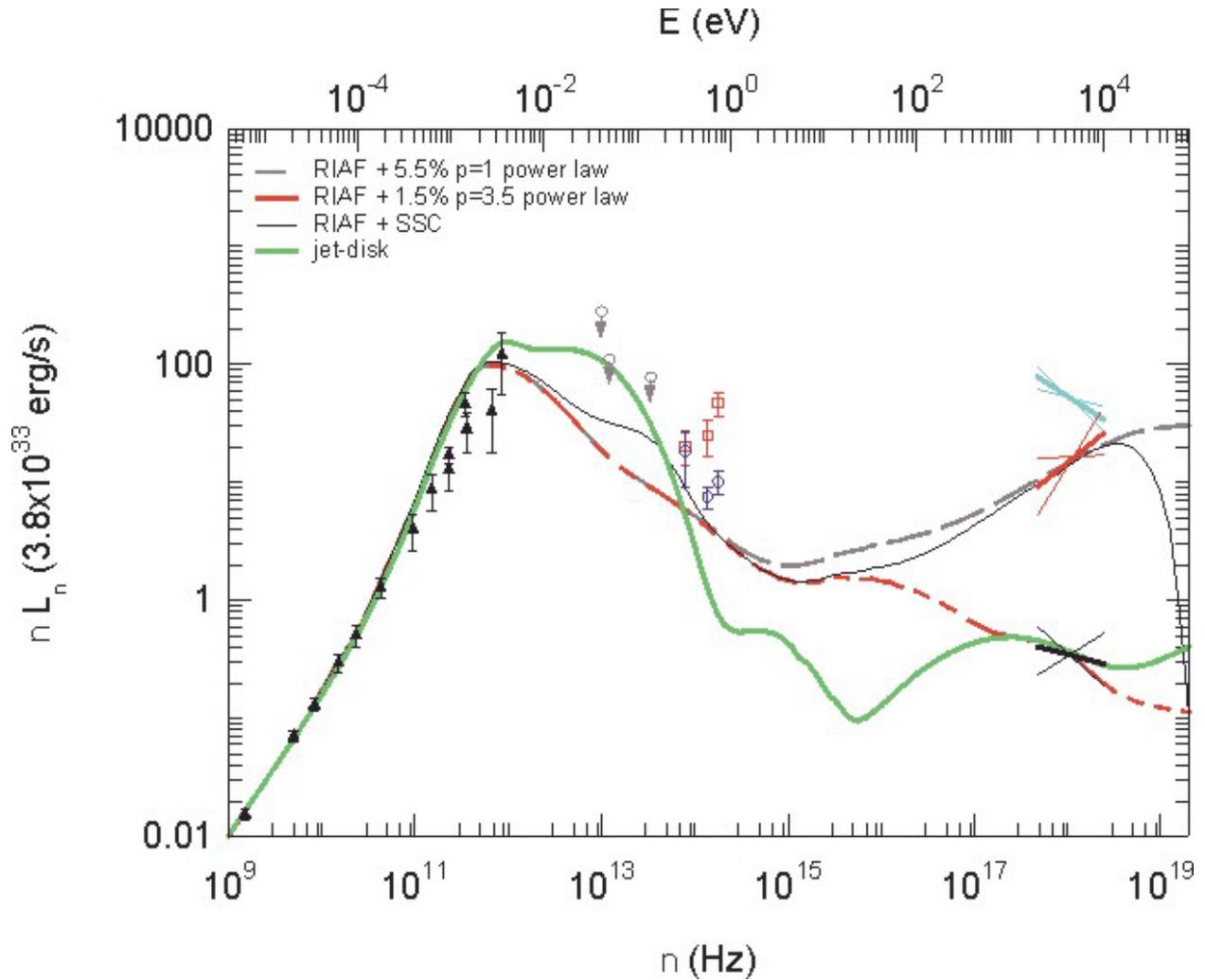

Figure 3 Spectral energy distribution emission from SgrA*. Plotted are the extinction and absorption corrected, band luminosities $\nu L_\nu$ (energy emitted per logarithmic energy interval) as a function of frequency, or energy. The observed flux density is $S_\nu = L_\nu / 4\pi D^2$, where $D = 7.94$ kpc is the Galactic Centre distance. All error bars are ±1$\sigma$. Black triangles denote the quiescent radio spectrum of SgrA*[18,21]. Open grey circles denote various infrared upper limits from the literature[18]. The three X-ray data ranges are (from bottom to top) the quiescent state as determined with Chandra[6] (black), the autumn 2000 Chandra flare[19] (red) and the autumn 2002 XMM flare[20] (light blue). Open red squares with crosses mark the peak emission (minus quiescent emission) observed in the four flares (Table 1; note that the measurements were taken at different times). Open blue circles denote the dereddened H, $K_s$ and $L_0$ luminosities of the quiescent state, derived from the local background subtracted flux density of the point source at the position of SgrA*, thus eliminating the contribution from any extended diffuse light from the stellar cusp around the black hole. Values plotted are from Table 1 (average of 2002–03 in the case of $L_0$ band). The thick green solid curve is the jet-starved disk model[15]. The red long dash-short dash curve is a radiatively inefficient accretion flow (RIAF) model of the quiescent emission, where in addition to the thermal electron population of ref. 15, 1.5% of the electrons are in a non-thermal power-law energy spectrum of exponent $p = -3.5$ (ref. 17). The black thin solid curve is a RIAF model of the flares with 5.5% of the electrons in a power law of $p = -1$ (ref. 17). The long-dash blue curve is a RIAF flare model of the flares with a synchrotron-self Compton component[17].